\documentclass[letterpaper]{llncs}{}
\usepackage{makeidx}  % allows for indexgeneration
\usepackage[misc]{ifsym}
\usepackage{amsmath,graphicx}
\usepackage{multirow}

\DeclareMathOperator{\Tr}{Tr}
\newcommand{\rpm}{\raisebox{.2ex}{$\scriptstyle\pm$}}
\begin{document}
\frontmatter          % for the preliminaries
\pagestyle{headings}  % switches on printing of running heads

\mainmatter              % start of the contributions
\title{A Generative-Discriminative Basis Learning Framework to Predict Clinical Severity from Resting State Functional MRI Data }
\titlerunning{Matrix Factorisation based Network Connectomics for ASD}  % abbreviated title (for running head)
\author{Niharika Shimona D'Souza\inst{1}\textsuperscript{*}
\and Mary Beth Nebel \inst{2} \inst{3} \and Nicholas Wymbs \inst{2} \inst{3} \and Stewart Mostofsky \inst{2}\inst{3} \inst{4} \and Archana Venkataraman\inst{1} }
%

%\author{***}
\authorrunning{Niharika Shimona D'Souza et al.} % abbreviated author list (for running head)
\institute{Dept. of Electrical and Computer Engineering, Johns Hopkins University, USA
\and
Center for Neurodevelopmental Medicine \& Research, Kennedy Krieger Institute,
\and
Dept. of Neurology, Johns Hopkins School of Medicine, USA
\and
Dept. of Pediatrics, Johns Hopkins School of Medicine, USA}

%\institute{***}

\maketitle              % typeset the title of the contribution
\begin{abstract}
We propose a matrix factorization technique that decomposes the resting state fMRI (rs-fMRI) correlation matrices for a patient population into a sparse set of representative subnetworks, as modeled by rank one outer products. The subnetworks are combined using patient specific non-negative coefficients; these coefficients are also used to model, and subsequently predict the clinical severity of a given patient via a linear regression. Our generative-discriminative framework is able to exploit the structure of rs-fMRI correlation matrices to capture group level effects, while simultaneously accounting for patient variability. We employ ten fold cross validation to demonstrate the predictive power of our model on a cohort of fifty eight patients diagnosed with Autism Spectrum Disorder. Our method outperforms classical semi-supervised frameworks, which perform dimensionality reduction on the correlation features followed by non-linear regression to predict the clinical scores.
%\keywords{Connectomics, Matrix Factorization, Basis Learning}
\end{abstract}
\section{Introduction}
Resting state fMRI (rs-fMRI) allows us to assess brain activity and localize critical functions through steady state patterns of co-activation \cite{fox2007spontaneous}. Building predictive models at the patient level remains an open challenge due to the high data dimensionality and to the considerable inter-subject variability. Predictive analysis methods usually follow a two step procedure. First, feature selection is applied to the raw correlation values; examples include graph theoretic measures, statistical measures and embedding features obtained from unsupervised learning techniques such as PCA, k-PCA or ICA \cite{murphy2012machine}. As a second step, conventional regression techniques such as Random Forests or Support Vector Regression are applied to the feature space representation to predict the clinical severity. These strategies adequately capture the group-averaged functional connectivity across the cohort but fail to account for individual variability. Consequently, the generalization power of these techniques is limited.
\par The recent success of Bayesian \cite{venkataraman2016bayesian} and dictionary learning \cite{batmanghelich2012generative} studies on rs-fMRI data is largely based on their ability to simultaneously model the patient and group level information. \cite{batmanghelich2012generative} introduces a basis learning framework for patient subtype classification, which reduces the dimensionality of T$1$ MR voxel based morphometry data while preserving the anatomical interpretability. \cite{eavani2015identifying} introduces a correlation matrix decomposition strategy, where multiple rank one matrix outer products capturing the underlying `generative' basis are combined using patient specific coefficients. The sparse basis networks identify meaningful co-activation patterns common to all the patients, and the coefficients model the patient variability. Our main contribution lies in exploiting the `discriminative' nature of rs-fMRI correlation matrices. We estimate the clinical severity of every patient by constructing a regression model which maps the behavioral scores to the functional data space. We jointly optimize for each of the hidden variables in the model, i.e. the basis, coefficients and regression weights. We refine the validation process by quantifying the model generalizability in terms of the regression performance on unseen data, as opposed to the correlation fit measure in \cite{eavani2015identifying}. Hence, our framework is less prone to overfitting. 
\par We validate our framework on a population study of Autism Spectrum Disorder (ASD). Patient variability manifests as a spectrum of impairments, typically quantified by a ``behavioral score" of clinical severity obtained from an expert assessment. Identifying sub-networks predictive of ASD severity is the key link to understanding the social and behavioral implications of the disorder. Our inclusion of behavioral data into the optimization framework guides the identification of representative networks specific to resting state ASD characterization.
\section{A Joint Model for Connectomics and Clinical Severity}Let $N$ be the number of patients and $M$ be the number of regions in our brain parcellation. We decompose the patient correlation matrices $\mathbf{\Gamma}_{n} \in \mathcal{R}^{M \times M}$ into a non-negative combination of a $K$ basis subnetworks {$\mathbf{b}_{k}\mathbf{b}_{k}^{T}$}. The sparse vector $\mathbf{b}_{k}$ indicates the relative contribution of each brain region to network $k$. The vector $\mathbf{c}_{n}$ denotes the non-negative contribution of each subnetwork for patient $n$. The coefficients $\mathbf{c}_{n} $ are subsequently used to model the clinical severity score $y_{n}$ via the regression weight vector $\mathbf{w} \in \mathcal{R}^{K}$. We concatenate the subnetworks into a basis matrix $\mathbf{B} \in \mathcal{R}^{M \times K}$, the coefficients into the matrix $\mathbf{C} \in \mathcal{R}^{K \times N}$, and the scores into a vector $\mathbf{y} \in \mathcal{R}^{N}$. Our combined objective can be written as follows:
\begin{eqnarray}
\mathcal{J}(\mathbf{B},\mathbf{C},\mathbf{w}) = {\sum_{n}}{\vert\vert{\mathbf{\Gamma}_{n}- \mathbf{B} {\textbf{diag}(\mathbf{c}_{n})} \mathbf{B}^{T}}\vert\vert}_{F}^{2} + \gamma{\vert\vert{\mathbf{\mathbf{y}}-\mathbf{C}^{T}\mathbf{w}}\vert\vert}_{2}^{2}  \ \ \  s.t. \ \ \mathbf{c}_{nk} \geq 0 ,\ \ \ 
\label{eqn1:Equation1}
\end{eqnarray}
Here, $\gamma$ is the tradeoff between the behavioral and functional data terms, and $\mathbf{diag}(\mathbf{c}_{n})$ is a matrix with the elements of $\mathbf{c}_{n}$ on its leading diagonal, and off diagonal elements as $0$. We impose an $\ell_{1}$ penalty upon the matrix $\mathbf{B}$ in order to recover a sparse set of subnetworks. Since the objective in Eq. (\ref{eqn1:Equation1}) is ill posed, we add quadratic penalty terms on $\mathbf{C}$ and $\mathbf{w}$ which act as regularizers.
\begin{equation}
\lambda_{1}{\vert\vert{\mathbf{B}}\vert\vert}_{1} + \lambda_{2}{\vert\vert{\mathbf{C}}\vert\vert}_{2}^{2} + \lambda_{3}{\vert\vert{\mathbf{w}}\vert\vert}_{2}^{2}.
\label{eqn2:Equation2}
\end{equation}
Eq. (\ref{eqn2:Equation2}) is added to the overall objective with $\lambda_{1}$, $\lambda_{2}$ and $\lambda_{3}$ being the sparsity, norm penalty on $\mathbf{C}$, and the penalty on $\mathbf{w}$ respectively.
\subsection{Optimization Strategy}
\par We employ a fixed point alternating minimization strategy to optimize $\mathbf{B}$, $\mathbf{C}$ and $\mathbf{w}$. At every iteration, the optimal solution for one variable is calculated assuming the other variables are held constant. Proximal gradient descent \cite{parikh2014proximal} is an effective strategy of optimizing a non-differentiable sparsity penalty such as the one in Eq. (\ref{eqn2:Equation2}), when the supporting terms are convex in the variable of interest. However, the expansion of the first Frobenius norm gives rise to non-convex bi-quadratic terms in $\mathbf{B}$, which prevents us from directly computing a proximal solution. Therefore, we introduce $N$ constraints of the form $\mathbf{D}_{n} = \mathbf{B}\mathbf{diag}(\mathbf{c}_{n})$, which are enforced by an Augmented Lagrangian penalty:
\begin{multline}
\mathcal{J}(\mathbf{B},\mathbf{C},\mathbf{w},\mathbf{D}_{n},\mathbf{\Lambda}_{n}) = {\sum_{n}}{\vert\vert{\mathbf{\Gamma}_{n}-\mathbf{D}_{n}\mathbf{B}^{T}}\vert\vert}_{F}^{2} +  \sum_{n}{\Tr{\left[{\mathbf{\Lambda}_{n}^{T}({\mathbf{D}_{n}-\mathbf{B}\mathbf{diag}(\mathbf{c}_{n})})}\right]}}   \\ + \sum_{n}{{\frac{1}{2}}{\vert\vert{\mathbf{D}_{n}-\mathbf{B}\mathbf{diag}(\mathbf{c}_{n})}\vert\vert}_{F}^{2}} + \gamma{\vert\vert{\mathbf{y}-\mathbf{C}^{T}\mathbf{w}}\vert\vert}_{2}^{2} \ \  s.t. \ \  \mathbf{c}_{nk} \geq 0
\label{eqn3:Equation3}
\end{multline} where, each $\mathbf{\Lambda}_{n}$ is a matrix of Lagrangians and each of the supporting Frobenius norm terms are regularizers on the Lagrangian constraints. The objective in \ Eq. (\ref{eqn3:Equation3}) is convex in $\mathbf{B}$ and the set $\{\mathbf{D}_{n}\}$ separately. Our optimization begins by randomly initializing $\mathbf{B}$, $\mathbf{C}$ and $\mathbf{w}$ and setting $\mathbf{D}_{n} = \mathbf{B}\mathbf{diag}(\mathbf{c}_{n})$ and ${\mathbf{\Lambda}_{n}} = \mathbf{0}$. We then iterate through the following four steps until global convergence.
\paragraph{\textbf{Step 1 - Optimizing B via Proximal Gradient Descent.}} Given the fixed learning rate parameter $t$, the proximal update for $\mathbf{B}$ is:
\begin{multline}
\mathbf{B}^{k+1} = \mathbf{sgn}(\mathbf{X}).^{*}(\mathbf{max}(\vert{\mathbf{X}}\vert-t,\mathbf{0})) \ \ s.t. \ \ \mathbf{X} = \mathbf{B}^{k} - (t/\lambda_{1})\frac{\partial \mathcal{J}}{\partial \mathbf{B}}
\end{multline}
The derivative of $\mathcal{J}$ with respect to $\mathbf{B}$, where $\mathbf{V}_{n} = \textbf{diag}(\mathbf{c}_{n})$, is computed as:
\begin{equation}
\frac{\partial \mathcal{J}}{\partial \mathbf{B}} = \sum_{n}\left[{{2\left[{\mathbf{B}\mathbf{D}_{n}^{T}\mathbf{D}_{n}-\mathbf{\Gamma}_{n}\mathbf{D}_{n}}\right]-\mathbf{D}_{n}\mathbf{V}_{n}+\mathbf{B}\mathbf{V}_{n}^{2}-\mathbf{\Lambda}_{n}\mathbf{V}_{n}}}\right]
\end{equation}
As seen, the non-smoothness of the ${\vert\vert{\mathbf{B}}\vert\vert}_{1}$ penalty is handled by performing iterative shrinkage thresholding applied on a locally smooth quadratic model.
\paragraph{\textbf{Step 2 - Optimizing C using Quadratic Programming.}}
The objective is quadratic in $\mathbf{C}$ when $\mathbf{B}$ and $\mathbf{w}$ are held constant. Furthermore, the $\mathbf{diag}(\mathbf{c}_{n})$ term decouples the updates for $\mathbf{c}_{n}$ across patients. Hence, we use $N$ quadratic programs of the form below to solve for the vectors $\{\mathbf{c}_{n}\}$ : 
\begin{equation}
\frac{1}{2}{\mathbf{c}_{n}^{T}\mathbf{H}_{n}\mathbf{c}_{n}} + \mathbf{f}_{n}^{T}\mathbf{c}_{n} \ \ s.t. \ \ \mathbf{A}_{n}\mathbf{c}_{n} \geq \mathbf{b}_{n}
\end{equation}
The Quadratic Programming parameters for our problem are given by:
\begin{eqnarray*}
\mathbf{H}_{n} = \mathbf{diag}(\mathbf{B}^{T}\mathbf{B}) + 2\gamma{\mathbf{w}\mathbf{w}^{T}}+ 2\lambda_{2}\mathcal{I}_{K} \ \ \ \ \ \ \ \ \ \ \ \ \ \\ \mathbf{f}_{n} = -\mathbf{diag}(\mathbf{D}_{n}^{T}\mathbf{B}) -\mathbf{diag}(\mathbf{\Lambda}_{n}^{T}\mathbf{B}) -2\gamma y_n\mathbf{w}; \ \ \ \ \mathbf{A}_{n} = -\mathcal{I}_{K} \ \ \ \mathbf{b}_{n} = \mathbf{0}
\end{eqnarray*}
This strategy helps us find the globally optimal solutions for $\mathbf{c}_{n}$, as projected onto the $K$ dimensional space of positive real numbers.
\paragraph{\textbf{Step 3 - Closed Form Update for $\mathbf{w}$.}} The global minimizer of $\mathbf{w}$ computed at the first order stationary point can be expressed as:
\begin{equation}
\mathbf{w} = (\mathbf{C}\mathbf{C}^{T}+ \frac{\lambda_{3}}{\gamma} \mathcal{I}_{K})^{-1}(\mathbf{C} \mathbf{y})
\end{equation}
\paragraph{\textbf{Step 4 - Optimizing the Constraint Variables $\mathbf{D}_{n}$ and $\mathbf{\Lambda}_{n}$.}}
Each of the primal variables $\{\mathbf{D}_{n}\}$ has a closed form solution given by: 
\begin{equation}
\mathbf{D}_{n} = (\mathbf{diag}(\mathbf{c}_{n})\mathbf{B}^{T}+ 2\mathbf{\Gamma}_{n}\mathbf{B} - \mathbf{\Lambda}_{n})(\mathcal{I}_{K}+2\mathbf{B}^{T}\mathbf{B})^{-1}
\end{equation}
In contrast, we update the dual variables $\{\mathbf{\Lambda}_{n}\}$ using gradient ascent:  
\begin{equation}
\mathbf{\Lambda}_{n}^{k+1} = \mathbf{\Lambda}_{n}^{k} + \eta_{k}(\mathbf{D}_{n}-\mathbf{B}\mathbf{diag}(\mathbf{c}_{n}))
\end{equation}
The updates for $\mathbf{D}_{n}$ and $\mathbf{\Lambda}_{n}$ ensure that the proximal constraints are satisfied with increasing certainty at each iteration. The learning rate parameter $\eta_{k}$ for the gradient ascent step  of the augmented Lagrangian is chosen to guarantee sufficient decrease for every iteration of alternating minimization. In practice, we initialize this value to $0.001$, and scale it by $0.75$ at each iteration. 
\par In all of  our derivations, $\Tr(\mathbf{M})$ is the trace operator and gives the sum of the diagonal elements of a matrix $\mathbf{M}$, and $\mathcal{I}_{K}$ is the $K \times K$ identity matrix. 
\subsection{Predicting Symptom Severity:} We use cross validation to evaluate the predictive power of our model. Specifically, we compute the optimal $\{\mathbf{B}^{\star},\mathbf{w}^{\star}\}$ based on the training dataset. We can then estimate the coefficients $\mathbf{c}_{test}$ for a new patient by re-solving the quadratic program in \textbf{Step} $\mathbf{2}$ using the previously computed $\{\mathbf{B}^{\star},\mathbf{w}^{\star}\}$. Notice that we must set the data term $\gamma{\vert\vert{\mathbf{C}^{T}\mathbf{w}-\mathbf{y}}\vert\vert}_{2}^{2}$ to $0$ in the testing experiments, since the severity $y_{test}$ is unknown. Also, we assume that the constraint $\mathbf{D}_{test} = \mathbf{B}^{\star}\mathbf{diag}(\mathbf{c}_{test})$ is satisfied exactly for the conditions of the proximal operator to hold. Finally, $y_{test} = \mathbf{c}_{test}^{T}\mathbf{w}^{\star}$ is the estimate of the behavioral score for the unseen test patient.
\subsection{Baseline Comparison Methods}
We compare our algorithm with a standard machine learning pipeline to predict the target severity score. We first perform dimensionality reduction to concentrate the $\frac{M \times (M-1)}{2}$ rs-fMRI correlation pairs into a small number of basis elements. Then, we construct a non-linear regression model to predict clinical severity. We consider two dimensionality reduction/regression combinations: 
\begin{itemize}
\item[1]{Principal Component Analysis on the correlation coefficients followed by a Random Forest Regression on the projected features}
\item[2]{Kernel Principal Component Analysis on the correlation coefficients followed by a Random Forest Regression on the embedding features}
\end{itemize}
\begin{center}
\begin{figure}[!t]
 \begin{minipage}[t]{1\linewidth}
 \centering 
 \centerline{\includegraphics[scale=0.23]{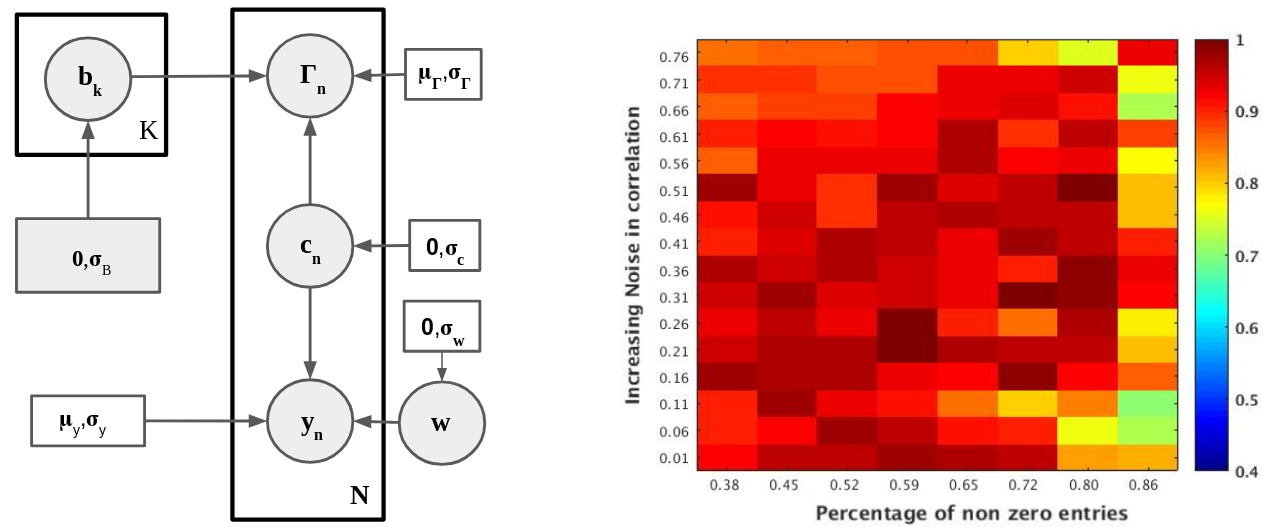}}
 \medskip
 \end{minipage}
 \caption{\footnotesize{(a) The graphical model from which we generate synthetic data (b) The network recovery performance of our algorithm for varying levels of sparsity and correlation matrix noise variance. For our experiments, we fix the rest of the parameters of the model at $\mathbf{\sigma}_{\mathbf{B}} = 0.2$, $\mathbf{\mu}_{\mathbf{\Gamma}_{n}} = \mathbf{B}\mathbf{diag}(\mathbf{c}_{n})\mathbf{B}^{T}$, $\mathbf{\sigma}_{\mathbf{c}} =0.1$, $\mathbf{\mu}_{\mathbf{y}}=\mathbf{c}_{n}^{T}{\mathbf{w}}$, $\mathbf{\sigma}_{\mathbf{y}}=0.2$ and $\mathbf{\sigma}_{\mathbf{w}}=0.1$}}
\label{Figure:1}
\end{figure}
\end{center}
\section{Experimental Results}
\paragraph{\textbf{Evaluating Robustness on Synthetic Data.}} Our optimization problem in Eq. (\ref{eqn1:Equation1}) suggests an underlying graphical model, depicted in Fig. \ref{Figure:1}(a). Notice that the $\ell_{1}$ penalty on $\mathbf{B}$ translates to a Laplacian prior with $\mathbf{\sigma}_{B}$ controlling the potentially overlapping level of sparsity. In contrast, the $\ell_{2}$ penalties translate into Gaussian distributions, with the mean corresponding to the $\ell_{2}$ argument and the variance related to the regularization parameters. We use this model to sample the correlation matrices $\{{\mathbf{\Gamma}}_{n}\}$ and the behavioral scores $\{y_{n}\}$, and then infer the latent networks generating the data. Fig. \ref{Figure:1}(b) indicates the performance of network recovery from our algorithm. We quantify the peformance in terms of average inner-product similarity between recovered networks and generating networks, both normalized to unit norm. The number of generating and recovery networks is chosen to be $4$. Unsurprisingly, increasing the overlap in the sparsity patterns across networks and increasing the noise in the correlation estimates worsens the recovery performance. However, our optimization procedure is robust in the noise regime estimated from our real-world rs-fMRI correlation matrices ($0.01-0.2$) and for recovered sparsity levels ($0.1-0.4$). The experiment also helps us identify stable parameter settings for the next section.
\paragraph{\textbf{rs-fMRI Dataset and Preprocessing.}} We evaluate our method on a cohort of $58$ children with high-functioning ASD (Age: $10.06 \rpm 1.26$, IQ: $110\rpm 14.03$). We acquired rs-fMRI scans on a Phillips $3$T Achieva scanner using a single-shot, partially parallel gradient-recalled EPI sequence (TR/TE $=2500/30$ms, flip angle $=70\deg$, res $=3.05\times3.15\times3$mm, $128$ or $156$ time samples).
\par Rs-fMRI preprocessing \cite{venkataraman2016bayesian} consisted of slice time correction, rigid body realignment, and normalization to the EPI version of the MNI template using SPM. We use a CompCorr  strategy to remove the spatially coherent noise from the white matter, ventricles, and six rigid body realignment parameters. We then spatially smoothed the data ($6$mm FWHM Gaussian kernel) and bandpass filtered the time series ($0.01-0.1$Hz). We use the Automatic Anatomical Labeling (AAL) atlas to define $116$ cortical, subcortical and cerebellar regions. Empirically, we observed a consistent noise component having nearly constant contribution from all the brain regions and low predictive power. Consequently, we subtract out the contribution of the first eigenvector from the correlation matrices and used the residuals $\{\mathbf{\Gamma}_{n}\}$ as inputs for all the methods.
\par We consider two measures of clinical severity: Autism Diagnostic Observation Schedule (ADOS) total raw score \cite{payakachat2012autism}, which captures the social  and communicative interaction deficits of the patient along with repetitive behaviors (dynamic range: $0-30$), and the Social Responsiveness Scale (SRS) total raw score \cite{payakachat2012autism} which characterizes social responsiveness (dynamic range: $70-200$).
\begin{table}[!b]
\centering
\caption{\footnotesize{Performance evaluation using \textbf{root median square error (rMSE)} \& $\mathbf{R^2}$ fit, both for testing \& training. Lower MSE \& higher $R^2$ score indicate better performance.}}
\small{
\begin{tabular}{|c| c| c| c| c| c|} 
\hline 
 \textbf{Score}& \textbf{ Metric } & \textbf{ Our Method } & \textbf{ PCA + RF Reg } & \textbf{ k-PCA + RF Reg } \\ [0.1ex] 
\hline 
\multirow{4}{4em}{ADOS} & rMSE train & \textbf{0.088}& 1.07 & 1.017 \\
 & $R^2$ train & \textbf{0.99} & 0.94  & 0.96 \\
 & rMSE test & \textbf{2.53} & 2.93  & 2.70 \\
 & $R^2$ test & \textbf{0.096} & 0.031 & 0.01 \\
\hline
\multirow{4}{4em}{SRS} & rMSE train & \textbf{0.13} & 6.43 & 6.90 \\ 
 & $R^2$ train & \textbf{0.99} & 0.95 & 0.97 \\ 
 & rMSE test &  \textbf{13.26}& 20.51 & 20.30\\ 
 & $R^2$ test & \textbf{0.052} & 0.023 & 0.008\\
\hline  
\end{tabular}
}
\label{table:1}
\end{table}
\paragraph{\textbf{Predicting ASD Severity.}} We employ a ten fold cross validation strategy for each of the methods, whereby, we train the model on a $90\%$ data split and evaluate the performance on the unseen $10\%$ test data. We perform a grid search to find the optimal parameter setting for each method. Based on these results, we fix the regression tradeoff at $\gamma = 1$, and the three regularization parameters at $\{\lambda_1 = 40,\lambda_2 = 2, \lambda_3 = 1\}$ for SRS, and $\{\lambda_1 = 30,\lambda_2 = 0.2, \lambda_3 = 1\}$ for ADOS, and the learning rate at $t = 0.001$ for proximal gradient descent. The number of components was fixed at $15$ for PCA and $10$ for k-PCA. For k-PCA, we use an RBF kernel with the coefficient parameter varied between $0.01-10$. \par
\begin{figure}[!t]
 \begin{minipage}[t]{1\linewidth}
 \centerline{\includegraphics[scale=0.23]{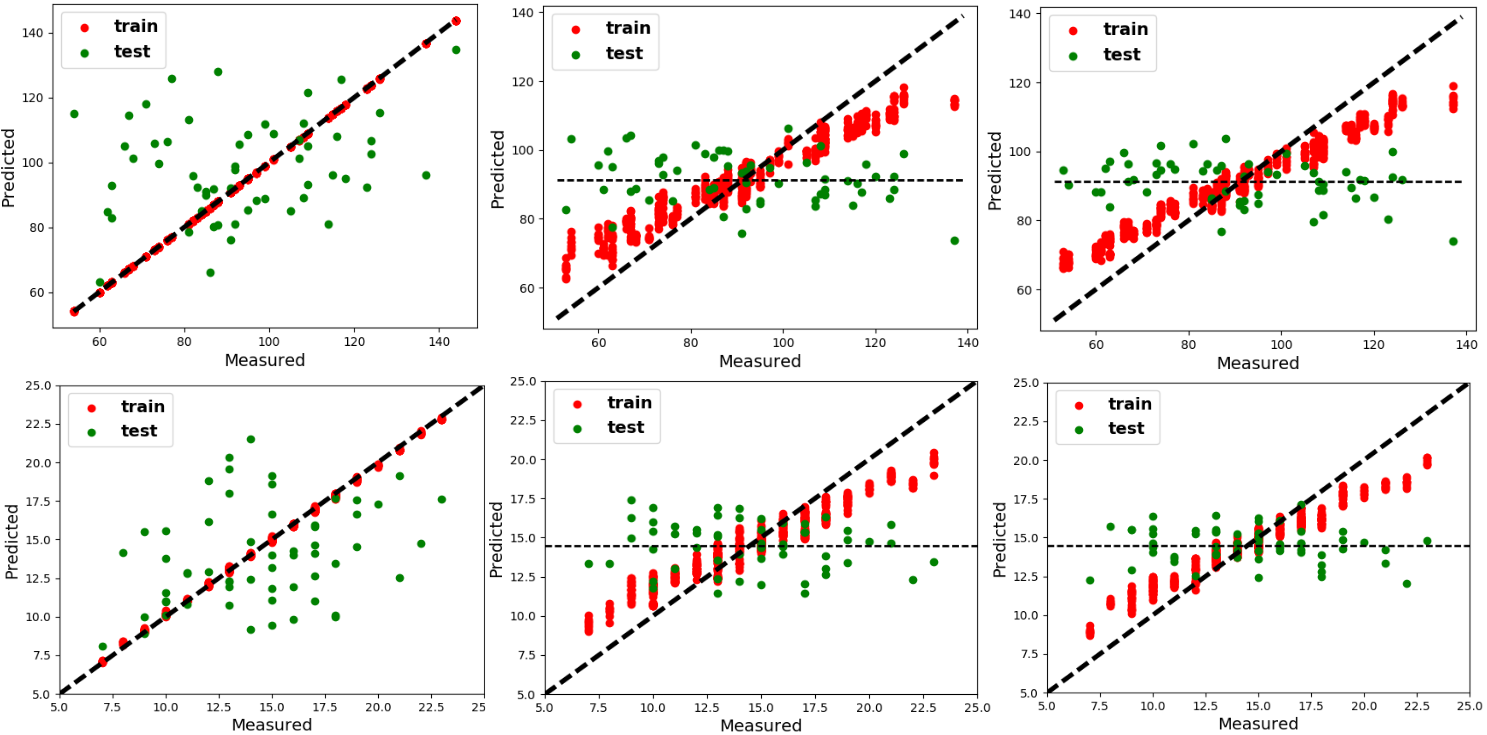}}
 \medskip 
 \end{minipage}
\caption{\footnotesize{Prediction performance of each method for ADOS (TR) \& SRS (BR) \textbf{Left:} Our Method ($K=8$) \textbf{Middle:} PCA ($comp=15$) \& RF Regression on the projected data \textbf{Right:} k-PCA ($comp=10$, rbf $C=0.1$) \& RF Regression on the embedding features. Red \& Green points correspond to the training \& testing performance respectively}}
 \label{Figure:2}
\end{figure}

\begin{figure}[t!]
 \begin{minipage}[t]{1\linewidth}
 \centerline{\includegraphics[scale=0.25]{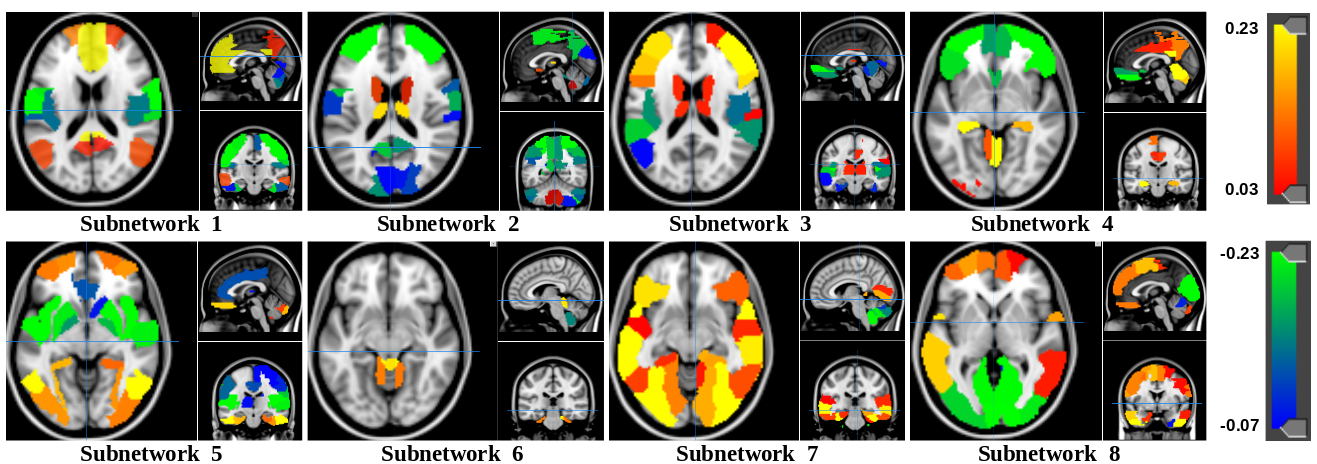}}
 \medskip 
 \end{minipage}
\caption{\footnotesize{Eight subnetworks identified by our model from ADOS prediction. The blue \& green regions are anticorrelated with the red \& orange regions for each subnetwork.}}
 \label{Figure:3}
\end{figure}

 As seen from the Fig. \ref{Figure:2}, the baseline methods have poor validation performance and track the mean value of the held out data (shown by the black line). In comparison, our method not only consistently fits the training set more faithfully, but also generalizes much better beyond the training data. The major shortcoming of the baseline data-driven analysis techniques is in their failure to identify representative patterns of behavior from the correlation features. In contrast, our basis learning technique exploits the underlying structure of the correlation matrices and leverages patient specific information to map the ASD behavioral space, thus improving the prediction performance. As reported in Table \ref{table:1}, our method quantitatively outperforms the baselines approaches, both in terms of the root median square error (rMSE) and the $R^2$ performance.
\paragraph{\textbf{Subnetwork Identification.}}
Fig. \ref{Figure:3} illustrates the basis subnetworks in $\mathbf{B}$ trained on the ADOS data. The colorbar indicates subnetwork contribution to the AAL regions. Regions storing negative values are anticorrelated with regions storing positive ones. Subnetwork $1$ includes competing i.e. anticorrelated contributions from regions of the default mode network (DMN) and somatomotor network (SMN). Abnormal connectivity within the DMN and SMN has been previously reported in ASD \cite{nebel2016intrinsic}. Additionally, subnetwork $5$ appears to be comprised of competing contributions from SMN regions and higher order visual processing areas in the occipital and temporal lobes, consistent with behavioral reports of reduced visual-motor integration in ASD. Subnetwork $2$ includes competing contributions from prefrontal and subcortical regions (mainly the thalamus, amygdala and hippocampus), that may be important for social-emotional regulation in ASD. Finally, subnetwork $3$ is comprised of competing contributions from the central executive control network and the insula, which is thought to be critical for switching between self-referential and goal-directed behavior \cite{sridharan2008critical}.
\section{Conclusion}
Unlike generic machine learning analysis, our matrix decomposition elegantly combines multimodal information from the imaging and behavioral domains. The key to our model is its ability to capture and learn from the structure of correlation matrices. Conventional analysis methods dramatically fall short of unifying the two data viewpoints reliably enough to implicate predictive functional patterns in the brain. Our joint optimization framework robustly identifies brain networks characterizing ASD and provides a key link to quantifying and interpreting the spectrum of manifestation of the disorder across a wide range of population. In the future, we will explore extensions of this model that jointly classify patients versus controls in addition to predicting symptom severity.\\ 
\\ \textbf{Acknowledgements.} This work was supported by the National Institute
of Mental Health (R01 MH085328-09, R01 MH078160-07, K01 MH109766 and R01 MH106564), the National Institute of Neurological Disorders and Stroke (R01 NS048527-08), and the Autism Speaks foundation.
{\bibliography{MyRefs}}
\bibliographystyle{splncs}
\end{document}